# Improved $H_2$–He and $H_2$-$H_2$ Collision-Induced Absorption Models and Application to Outer-Planet Atmospheres

Glenn S. Orton[1], Magnus Gustafsson[2], Leigh N. Fletcher[3], Michael T. Roman[3,4], and James A. Sinclair[1]

[1]Jet Propulsion Laboratory, California Institute of Technology,4800 Oak Grove Drive, Pasadena, CA 91109, USA

[2]Applied Physics, Division of Materials Science, Department of Engineering Science and ' Mathematics, Luleå University of Technology, 97187 Luleå, Sweden

[3]School of Physics and Astronomy, University of Leicester, University Road, Leicester, LE1 7RH, UK.

[4] Facultad de Ingeniera y Ciencias, Universidad Adolfo Ibáñez, Av. Diagonal las Torres 2640, Peñalolén, Santiago, Chile

(Dated: 2025 May 21)

## ABSTRACT

Using state-of-the-art *ab initio* interaction-induced dipole and potential-energy surfaces for hydrogen–helium ($H_2$–He) pairs, we compute the rototranslational collision-induced absorption coefficient at 40-400 K for frequencies covering 0-4000 cm$^{-1}$. The quantum mechanical scattering calculations account for the full anisotropic interaction potential, replacing the isotropic approximation. The absorption data are expected to be accurate with an uncertainty of 2% or better up to 2500 cm$^{-1}$. The uncertainty is slightly higher at the highest frequencies where the rototranslational absorption is largely obscured by the rovibrational band. Our improved agreement with measurements at 200-800 cm$^{-1}$ result from the improvement of the potential energy surface. The previously available rototranslational data set for $H_2$–$H_2$ pairs (Fletcher et al., *Astrophys. J. Supp.* **235**, 24 (2018)) is also extended up to 4000 cm$^{-1}$. In the rovibrational band previous isotropic potential calculations for $H_2$–He (Gustafsson et al., *J. Chem. Phys.* **113**, 3641 (2000)) and $H_2$–$H_2$ (Borysow, *Icarus* **92**, 273 (1992)) have been extended to complement the rototranslational data set. The absorption coefficients are tabulated for *ortho*-to-*para* ratios from normal-$H_2$ to pure *para*-$H_2$, as well as equilibrium-$H_2$, over 40-400 K . The effect of these updates are simulated for the cold atmosphere of Uranus and warmer atmosphere of Jupiter. They are equivalent to a brightness temperature difference of a fraction of a degree in the rototranslational region but up to 4 degrees in the rovibrational region. Our state-of-the-art modifications correct an otherwise +2% error in determining the He/$H_2$ ratio in Uranus from its spectrum alone.

    I.    **Introduction**

Accurate radiative-transfer modeling of outer-planet atmospheres requires careful consideration of the $H_2$-$H_2$ and $H_2$-He collision-induced absorption (CIA), as these two constituents dominate the compositions of Jupiter, Saturn, Uranus and Neptune. Besides being important in modeling the radiant energy control of the upper tropospheres and lower stratospheres of each of these planets, they are the only two major atmospheric constituents that are uniformly mixed in all of them. Thus, they are essential in retrievals of temperature, gas composition and aerosol properties in each of these atmospheres for mid- and far-infrared thermal emission, and for multiple scattering of reflected sunlight in the near-infrared. The increasing precision of observations made available by sophisticated facilities, such as the James Webb Space Telescope (JWST), call for a commensurate consideration for the basis of opacities that are used in these retrievals. A major step towards this level of precision from semi-empirical results (see Birnbaum et al.[1] and references therein), was made by Orton et al.[2], who improved a semi-empirical model for $H_2$-$H_2$ translational-rotational collision-induced absorption by adopting the results of ab-initio quantum-mechanical models, which provided a more trustworthy basis for mid- and far-infrared models than semi-empirical models[1]. This model was subsequently improved by repeating the calculations over a range of *para*-$H_2$ fractions for $H_2$-$H_2$ up to 2400 cm$^{-1}$ and including the role of $(H_2)_2$ dimers, as detailed by Fletcher et al.[3]. In the absence of in situ exploration, much of our understanding of planetary composition, dynamics, temporal variability, chemistry and clouds comes from inversions of infrared spectra; thus, we are refining the very foundations of all that work here, and providing a resource for the community as it embarks on analysis of JWST and related observations. For example, these improvements were particularly important for radiative-transfer modeling of the relatively cold atmospheres of Uranus and Neptune, as adopted by Orton et al.[4] and Roman et al.[5]

In this paper, we report further refinements and extensions of these models. First, we will describe a model for $H_2$-He CIA that is also based on ab-initio quantum mechanics, allowing a level of precision that is equal to or better than that of $H_2$-$H_2$ CIA. In particular, this enables accurate modeling of the far-infrared spectra of Uranus and Neptune, which is dominated by the $H_2$-related CIA from 18 μm to 3 mm and modulated by the relative abundances of $H_2$ and He. The He/H ratio is important for cosmogonic considerations as well as interior modeling. We will then describe an extension of the $H_2$-$H_2$ ab-initio quantum mechanical model to a higher-wavenumber region in the near infrared. This has become important in careful modeling of the JWST spectra of the outer planets, particularly in consideration of the depths to which aerosol absorption and scattering vs gaseous absorption dominates the spectrum. This is vital in order to attribute appropriate condensation levels for constituents, such as $CH_4$ and $H_2S$ (see Irwin et al.[6] and references therein), as well as providing the continuum absorption on top of which lie trace gaseous spectral features. Below, we first describe the basis of the absorption models. We then demonstrate how these revised absorption coefficients modify the $H_2$-He and $H_2$-$H_2$ CIA radiative-transfer calculations compared with previous models for the warmest and coldest atmospheres, Jupiter and Uranus, respectively.

## II. Calculations of collision-induced absorption

### A. $H_2$-He, rototranslational band

A method to compute binary collision-induced absorption for $H_2$–He was put forward and tested by Gustafsson et al.[7] It is based on quantum mechanical close-coupled (CC) scattering calculations[8], so that the full anisotropic interaction potential can be accounted for. Such calculations are computationally costly and highly sensitive to choices of numerous parameter values. In addition, they have to be repeated for many energies in order to obtain a Boltzmann-averaged, temperature-dependent absorption coefficient at each point of a dense frequency grid.

Under the approximation that the interaction-potential is independent of the orientation of the $H_2$ molecule, relative to the position of the He atom, the potential depends only on one variable: the distance from the atom to the molecule. Consequently, the couplings vanish in the coupled Schrödinger equation, and one is left with solving a set of uncoupled ordinary differential equations[9]. We will call this the isotropic potential approximation (IPA).

It was established by Gustafsson et al.[7] that the IPA absorption profiles for $H_2$–He mimic those computed with the full, anisotropic, potential. There are differences in absorption up to about 20% in magnitude (see Fig. 1), but the general shapes are quite similar. A machine-learning approach was explored to take advantage of this, and produced highly accurate absorption data on a dense frequency grid[10]. Here we present a simpler alternative to machine learning, in order to achieve the same goal. The approach is illustrated in Fig. 1 and works as follows. We consider the ratio of absorption coefficients using the anisotropic and isotropic potentials (CC/IPA) at each frequency. In the case of the 195-K data shown in Fig. 1 this ratio is a number that varies between about 0.92 to 1.13 over the spectral profile. Then we interpolate linearly to estimate the ratio in between the CC grid points. Finally, this ratio is multiplied by the IPA absorption on each IPA grid point to fill in with accurate absorption on a dense grid. We denote the result as the amended anisotropic (AA) data.

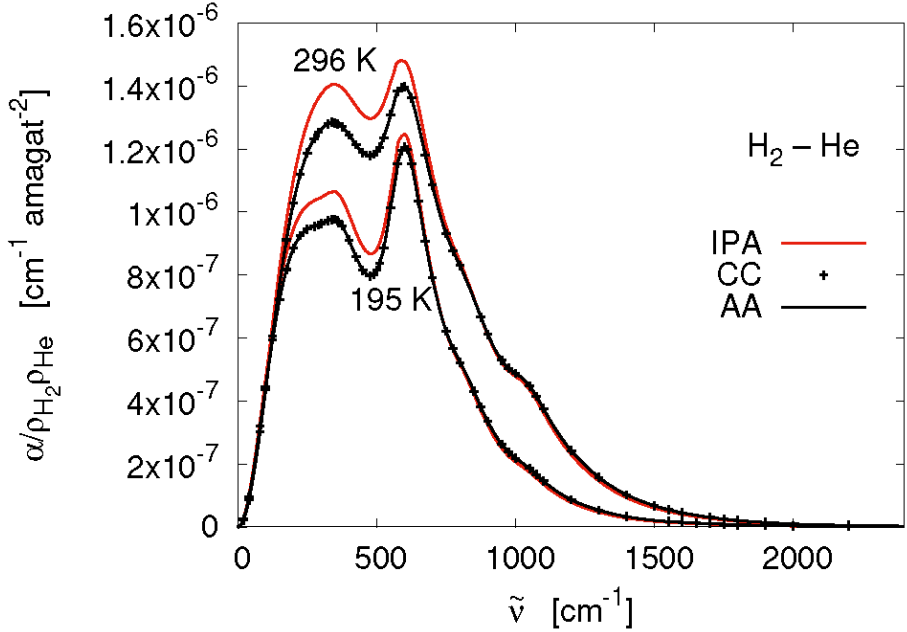

Figure 1. Binary collision-induced $H_2$–He absorption coefficient, α, normalized by the hydrogen and helium gas densities, as function of frequency in the rototranslational band, at the temperatures of 296 K (upper trace) and 195 K (lower trace). Equilibrium hydrogen, with respect to para-to-*ortho* distribution, is assumed at both temperatures. The red curves (IPA) are computed with the isotropic potential approximation[9]. The black crosses (CC) are computed with the close coupling approach[7], which accounts for the full anisotropic potential. The solid black curves (AA) represent the amended anisotropic data, obtained with the method outlined in section II A.

The calculations in this work, both CC and IPA, are performed using the Bakr, Smith, Patkowski (BSP) potential energy surface[11] and the interaction-induced dipole moment from Gustafsson et al.[7] To produce the data base we compute the IPA $H_2$–He absorption coefficient for

- 255 frequencies, from 0.1 cm$^{-1}$ to 4,000 cm$^{-1}$
- 10 temperatures, from 40 K to 400 K
- 11 *para*-$H_2$ to ortho-$H_2$ fractions, $f_p$ from 0.25 to 1.0, and equilibrium $H_2$ at each temperature

In order to compare with laboratory measurements, we have also done both CC and IPA calculations for 195 K and 296 K, which is also used for the illustration in Fig. 1. The CC absorption coefficient is computed on a sparser frequency grid with around 60 points, but for the same temperatures and *para*/*ortho*-fractions. As in Gustafsson et al.[7] the CC calculations are carried out using the COUPLE scattering program[11] with implementation based on Julienne[8]. The IPA calculations, on the other hand, are carried out with an in-house code that uses a Numerov algorithm for integration of the scattering Schrödinger equation. Details are given about the IPA and CC calculations in Appendices A and B, respectively.

In Figure 2 our amended anisotropic data at 195 K and 296 K are compared with experimental data[13,14,15]. We note that the agreement is good over the whole frequency range. It is an improvement from the previous CC calculation[7], which underestimated the absorption from about 200 cm$^{-1}$ to 800 cm$^{-1}$ that includes the S(0) and S(1) regions. We conclude that this is thanks to the BSP potential energy surface being more accurate than the Schaefer–Köhler (SK) potential[16], which was used in Gustafsson et al.[7] In Appendix B we elaborate on alternative potential energy surfaces and the difference between absorption obtained with the BSP and SK potentials.

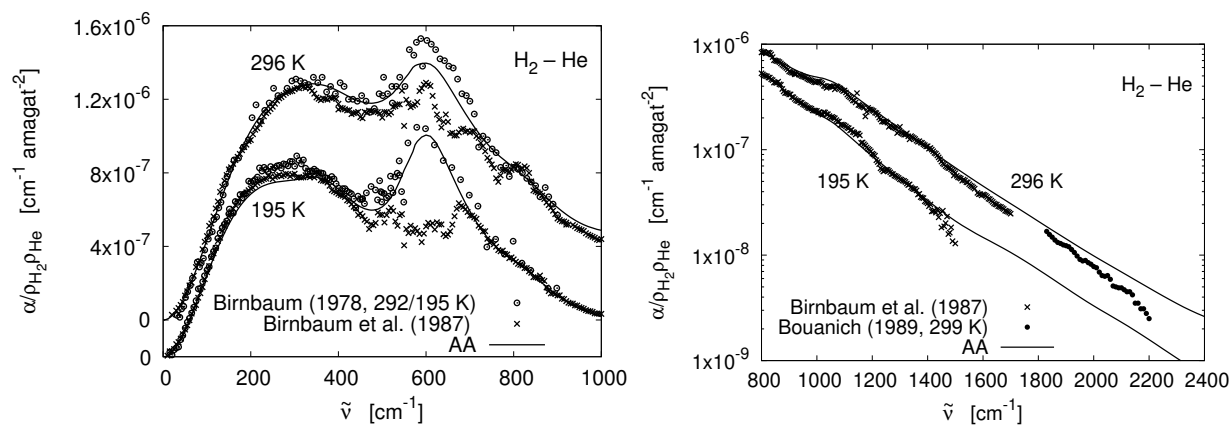

Figure 2. Binary collision-induced $H_2$–He absorption coefficient, α, normalized by the hydrogen and helium gas densities, as a function of frequency in the rototranslational band, at the temperatures of 296 K (upper trace) and 195 K (lower trace). Equilibrium hydrogen, with respect to para-to-*ortho* distribution, is considered at both temperatures. Solid curves represent our calculations according to the method outlined in section II A. Left panel: The data and model at the lower temperature are shifted downward by 2 x 10$^{-7}$ cm$^{-1}$ amg$^{-2}$ for the sake of clarity. For comparison we have included laboratory measurements by Birnbaum[13] at 292 K and 195 K and Birnbaum et al.[14] with 35% He in the gas mixture. Right panel: Semi-logarithmic plot including laboratory measurements from Birnbaum et al.[14], with 35% He in the gas mixture, and Bouanich et al.[15] taken at 299 K. Evidently, for Birnbaum et al.[14] there are some problems with the data around 600 cm$^{-1}$. Laboratory measurements of $H_2$-He CIA are challenging, among other things due to the fact that $H_2$-$H_2$ CIA has to be subtracted from the recorded absorption.

## B. $H_2$–He, fundamental band

Gustafsson et al.[7] also computed the $H_2$ rovibrational fundamental band absorption, i.e. with $H_2$ vibration going through transitions 0 → 1, peaking in the 3800 – 5000 cm$^{-1}$ range. This was done both accounting for the anisotropy of the interaction potential and in the IPA. Here, in order to obtain a data set towards shorter wavelengths, we have extended the IPA calculation of Gustafsson et al.[7] in frequency and temperature, range and grid density at the same temperatures and *para/ortho*-fractions as in section II A. We deem the IPA, rather than the costly CC calculation, to be sufficient for the $H_2$ fundamental spectral region in this work, as the CC result is closer to the IPA result in the $H_2$-He fundamental than it is in the $H_2$-He rototranslational absorption. The same interaction potential as in Gustafsson et al.[7] is used for these calculations, namely that of Ref. 16. The calculation covers the frequency range from 2000 cm$^{-1}$ to 8000 cm$^{-1}$, which, in absolute absorption, is safely smaller than the rototranslational and first overtone bands at the lower and upper end, respectively. Figure 3 shows a comparison of our results with the model of Abel et al.[17] and with the laboratory results of Vitali et al.[18] and Brodbeck et al.[19]. There is a consistent difference between these two sets of experimental results, but the systematic uncertainties Vitali et al.[18] associate with their measurements substantially overlap with uncertainties of the absorption spectra of Brodbeck et al.[19] in this region. The biggest differences between our model, the model of Abel et al.[16] and the mean of the experimental observations are at the 6% level. We note that our model has been optimized for the temperatures of the giant planets, rather than the warmer (200 – 9900 K) range considered by Abel et al.[17] (We note that 200 K is only achieved at pressures greater than 1 bar in outer-planet atmospheres. deeper than most infrared spectra sense.)

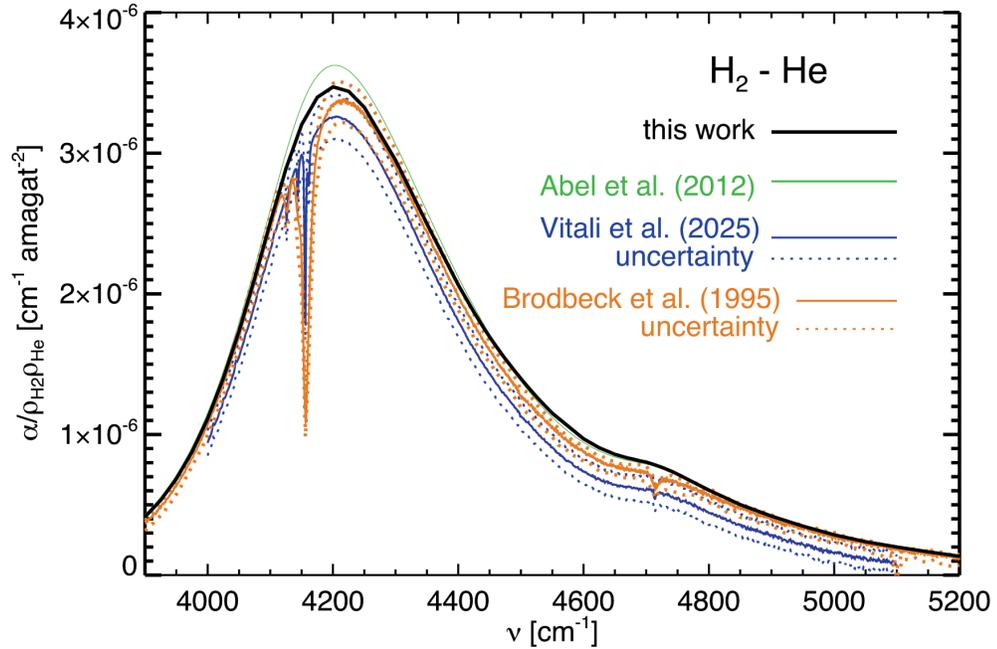

Figure 3. Binary collision-induced $H_2$–He absorption coefficient, $\alpha$, normalized by the hydrogen and helium gas densities, as function of frequency in the fundamental band, at 296 K. The *para*-to-*ortho* distribution, is representative of equilibrium $H_2$, just as in Fig. 2. Our model ("this work"), is compared with the model of Abel et al.[17] for T = 300K, and with the laboratory measurements of Vitali et al.[18] and Brodbeck et al.[19], both of which were made near 296 K. We note that neither our model nor that of Abel et al. considered the discrete, intercollisional interference dips present in the laboratory spectra.

### C. $H_2$–$H_2$, rototranslational band

The previously published data set for $H_2$–$H_2$ rototranslational collision-induced absorption[2,3] extends to 2400 cm$^{-1}$. In order to facilitate a smooth and reliable transition from the rototranslational band to available data sets for the fundamental band, we have computed the rototranslational absorption coefficient from 2500 cm$^{-1}$ to 4000 cm$^{-1}$. This was done in the same way as in Orton et al.[2] and Fletcher et al.[3], but suppressing bound dimer contributions, for the same temperatures and *para*-fractions as listed above in Section II A. The potential and dipole surfaces come from Schäfer & Köhler[20] and Meyer et al.[21], respectively. The calculations are done in the isotropic potential approximation using our in-house Numerov code. The corresponding parameter values are tabulated in Appendix C.

### D. $H_2$–$H_2$, fundamental band

Over three decades ago Borysow[22] computed the $H_2$–$H_2$ absorption in the fundamental band and composed a FORTRAN code to reproduce the spectrum from an analytical fit to the computed absorption at temperatures from 20 K to 300 K, for equilibrium and normal hydrogen. For our data base we have augmented that code so that absorption at arbitrary *para*-$H_2$ fractions, $f_p$, can be obtained. This is done according to the formulas presented in Appendix D.

## III. Outer-Planet Modeling

Here, we illustrate the differences between the currently used values for $H_2$-He and $H_2$-$H_2$ CIA coefficients (Borysow et al.[22], Borysow et al[24], respectively), and the revised values derived here. It is instructive to demonstrate the extent to which the revised absorption coefficient affects the radiative-transfer models for atmospheres in the outer solar system. We do this using the coldest and warmest atmospheres to which they would be applied, those of Uranus and Jupiter, respectively.

### A. $H_2$-He, rototranslational band

Figure 4 demonstrates differences between (i) the current model $H_2$-He rototranslational CIA spectrum, based on Borysow et al.[23], and (ii) this work for temperatures relevant to the atmospheres of Uranus and Jupiter. The temperature profile assumed for Uranus is the one derived from Spitzer Infrared Spectrometer (IRS) observations[4], and the temperature profile assumed for Jupiter is a smoothed version of the one retrieved by the Galileo Probe Atmospheric Structure Instrument[24]. The most notable differences from Borysow et al. result from our inclusion of dipole components 43 and 45, that create U transitions at 1622 and 2016 cm$^{-1}$, which were apparently omitted in their model. This effectively increases the opacity of the atmosphere in this region of weak absorption, somewhat closing a window into the deep region of the atmosphere.

Below wavenumber 150 cm$^{-1}$, the revised values for $H_2$-He CIA in both temperature regimes are lower than those of Borysow et al.[23] by up to 10%. For higher frequencies, differences are minimal up to approximately 500 cm$^{-1}$, after which the differences are in the opposite sense and rise to as high as 60 - 80% at the highest wavenumbers.

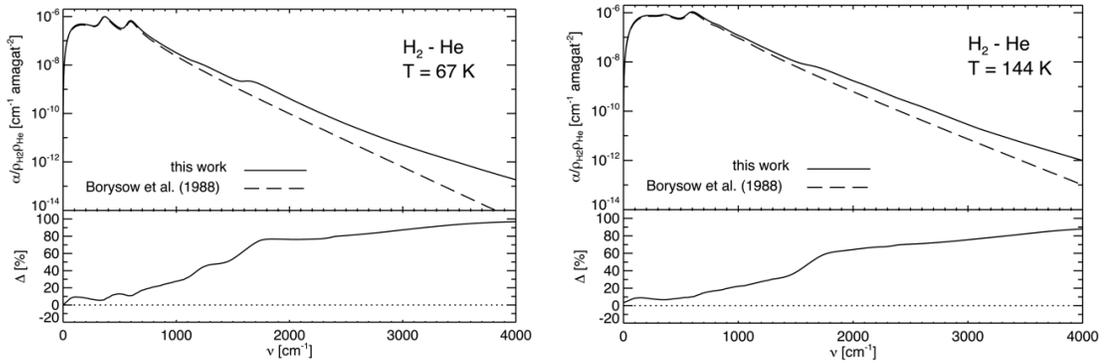

Figure 4. Left: (top) Absorption coefficient spectra for the rototranslational $H_2$-He collision-induced spectra at a temperature close to the mean brightness temperature of the upwelling radiance in this spectral region for Uranus. These are shown for the work in Borysow et al.[23] (dashed lines) and the work described in this paper (solid lines). (bottom) Spectral difference between them in percentage. Right: Similar to the left panel but for Jupiter. The temperatures chosen for this illustration are elements of the temperature grid used in the *ab initio* calculations, listed in the tables in the Appendices. Here and in Figs. 5 and 6, the curves are smoothed in frequency to suppress $H_2$ dimer contributions for clarity. For reference in these and in subsequent plots of residuals, zero is denoted by a dotted line.

Figures 5 and 6 show the effects of adopting our revised $H_2$-He CIA values in the atmospheres of Uranus and Jupiter, respectively, ignoring all other sources of opacity other than the $H_2$-$H_2$ and $H_2$-$CH_4$ CIA. These differences are illustrated both in terms of percentage differences in the upwelling radiance and in the equivalent brightness temperature. For Uranus, the model spectrum resulting from using our $H_2$-He CIA coefficients is about 7% lower than using those of Borysow et al.[23], centered in the 1500-2200 cm$^{-1}$ spectral range. This is equivalent to a brightness temperature that is 0.4 K lower in this range. For Jupiter, these differences are smaller, with our models being 1.5% lower in radiance or 0.3 K lower in brightness temperature in a similar spectral range.

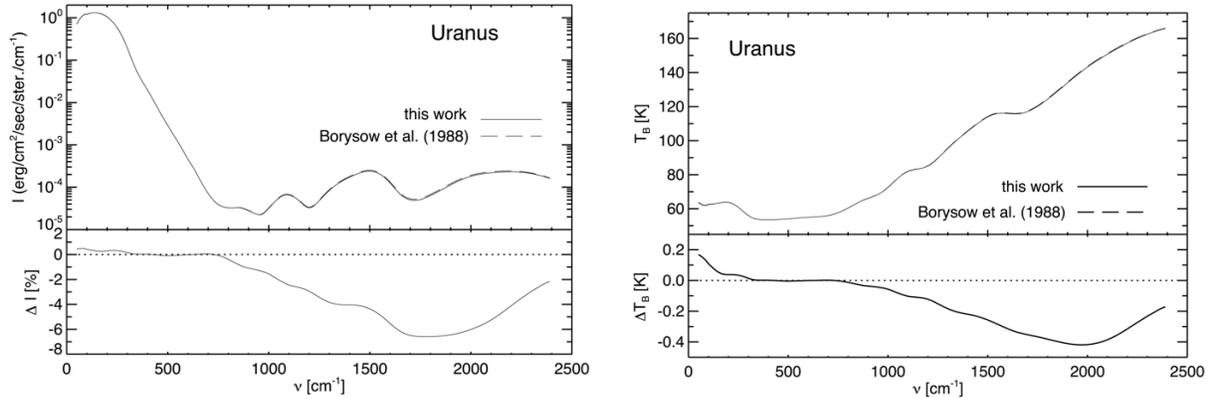

figure. 5. Left: (top) Models for the radiance of nadir thermal emission emerging from Uranus considering only the opacity provided by $H_2$-related collision-induced absorption, assuming equilibrium *para*-to-*ortho* fractions of $H_2$ at each level. The differences in the spectra emerge from using the model for $H_2$-He CIA based on Borysow et al.[23] (dashed line) and for the model presented here (solid line). (bottom) A spectrum of the percentage difference between radiance of the models. Right: Similar to the left panel but displaying the (top) radiance results in terms of equivalent brightness temperature and (bottom) brightness-temperature differences between the model spectra.

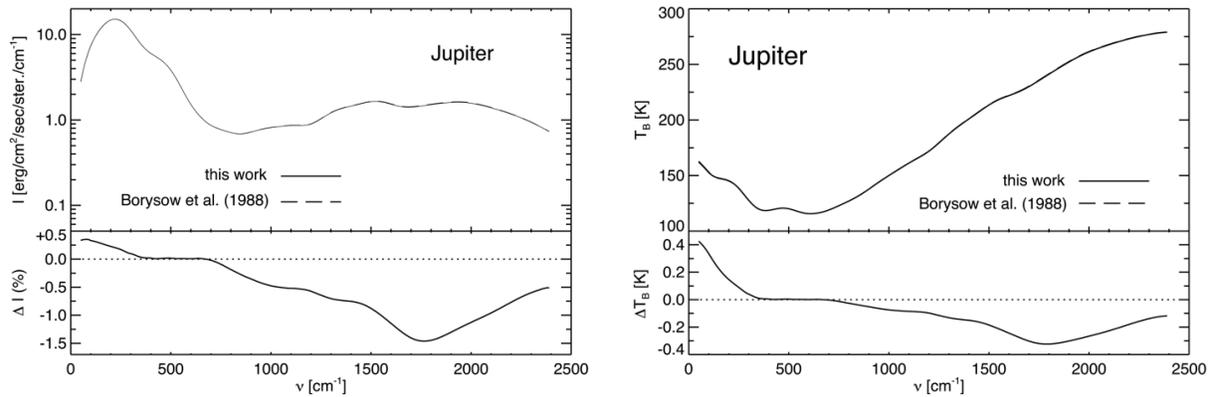

Figure 6. Similar to Fig. 5 but for the atmosphere of Jupiter.

**B. $H_2$-He, fundamental band**

Our revision of the $H_2$-He CIA model spectrum was extended to 8000 cm$^{-1}$, which encompasses the fundamental vibrational band. Figure 7 shows the comparison of our model with that of Borysow[24] using the same representative temperatures as in Figure 4 for Uranus and Jupiter. The largest differences are between our model and the calculations of Borysow[24] is around 5300 and 2700 cm$^{-1}$, for the U(0) and U(1) bands, respectively, which reach factor of ~3 higher than in our calculations. Those calculations were done with an ab initio interaction-induced dipole from Frommhold and Meyer[26]. In that work only three angular configurations of the $H_2$-He complex were considered, and the dipole was expanded in four angular components $\Lambda$ = 01, 21, 23, 45. The U(0) and U(1) features stem from the last dipole component (45). In the present work we use a more elaborate interaction-induced dipole surface Gustafsson et al.[7] where four angular configurations of $H_2$-He were considered, and the dipole is expanded in five angular components $\Lambda$ = 01, 21, 23, 43, 45. We conclude that this improvement has a rather drastic influence on the U(0) and U(1) transitions in the fundamental band at 5271 and 5695 cm$^{-1}$, respectively.

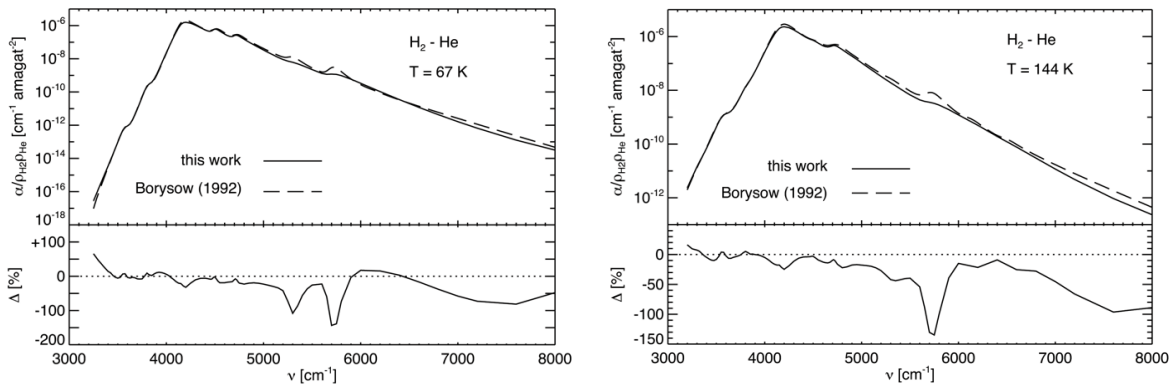

Figure. 7. Left: (top) Absorption coefficient spectra for the fundamental $H_2$-He collision-induced spectrum at a temperature relevant to Uranus from Borysow[24] (dashed lines) and for this work (solid lines). (bottom) Percent difference between the coefficients. Right: Similar to the left panel, but for a temperature more representative of Jupiter's atmosphere. The respective temperatures are the same as used in Fig. 4.

Figure 8 illustrates the differences in using the $H_2$-He CIA coefficients of Borysow[24] and our models in calculating the depth of reflected sunlight from the atmosphere of Uranus (left panel) and from the atmosphere of Jupiter (right panel). This region is dominated for both planets by reflected light rather than thermal emission. Because there is no consensus on a 'standard' cloud model for either planet (unlike a standard temperature profile, as used in Figs. 5 and 6), we have simply calculated the differences between the models for the depth at which the 2-way absorption from a perfectly reflecting layer reaches unity. For both planets the differences are small between 3000 and 6000 cm$^{-1}$. Between 2500 and 3000 cm$^{-1}$, our calculations show more absorption and a 30-60% lower pressure for our CIA models for Uranus and a smaller 20% difference for Jupiter. Between 6000 and 7500 cm$^{-1}$ our calculations for a Uranus model reach pressures 80% lower than those using the coefficients of Borysow[24]. For the same spectral region our calculations for a Jupiter model reach pressures about 30% lower than those using the coefficients of Borysow[24]. These are larger than the ~8% differences seen in the rototranslational spectrum around the overlapping wavenumbers.

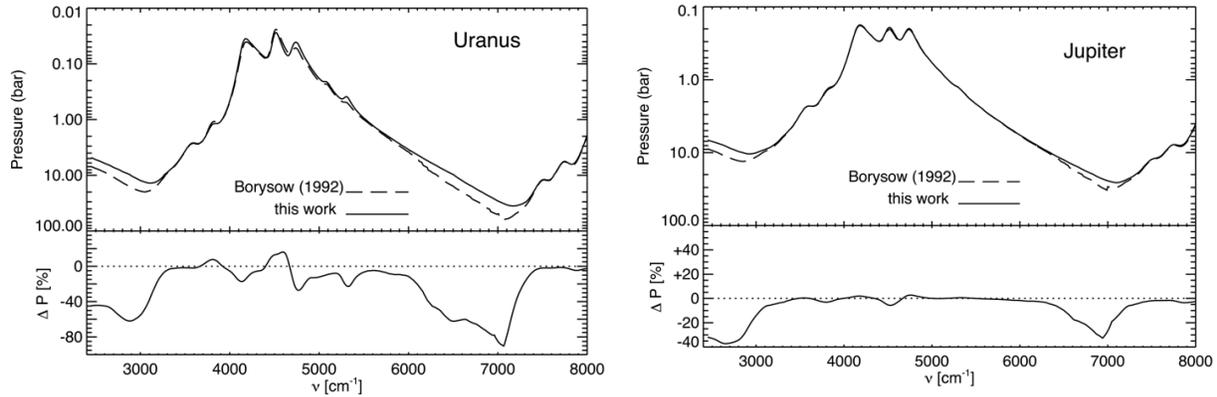

Figure. 8. Left: (top) Pressure in the atmosphere at which 2-way optical depth unity is reached in the atmosphere of Uranus. This effectively determines the maximum depth at which half of the sunlight in this spectral region is reflected back to the observer viewing from a nadir perspective from a perfectly reflecting cloud top at that level. The calculations considered only the fundamental $H_2$ collision-induced spectrum, including the $H_2$-He absorption coefficients from Borysow[24] (dashed line) and those described in this paper (solid line). (bottom) Percentage difference between the two models. Right: Similar to the left panel, except for the atmosphere of Jupiter. No other opacity sources were included in these calculations. The calculations shown in this figure do not include the overtone bands, which must be addressed in future work.

### C. $H_2$-$H_2$, rototranslational band

As a part of this study, we wanted to develop $H_2$-$H_2$ absorption coefficients that extended to higher frequency values (2500-4000 cm$^{-1}$) the calculations that formed the basis of Refs. 2 and 3 (apart from the contribution of dimers). Figure 9 illustrates the difference between these results and those of Borysow et al.[27], using the same two temperatures used for demonstrations of relevance to Uranus and Jupiter as in Figs. 3 and 6. Figure 8 shows that there is a substantial difference between the two at wavelengths between 2500 and 4000 cm-1, with the CIA $H_2$-$H_2$ coefficients derived here being 50-60% higher than those of Borysow et al.[27]

Figure 10 shows the difference between the spectra of Uranus using the coefficients derived by this study and those of Borysow et al.[27] that were illustrated for two discrete temperatures in Figure 9. For the atmosphere of Uranus, the higher opacity of the coefficients derived here, produce a radiance that is as much as 50% lower than one using the coefficients of Borysow et al.[27], equivalent to a brightness-temperature difference that is 3.7 K lower. These differences rapidly diminish with higher frequencies. For Jupiter, Figure 10 shows a qualitatively similar behavior, but with a radiance that is 17% lower than one using the coefficients of Borysow et al.[27], equivalent to brightness-temperature difference that is 3.5K lower.

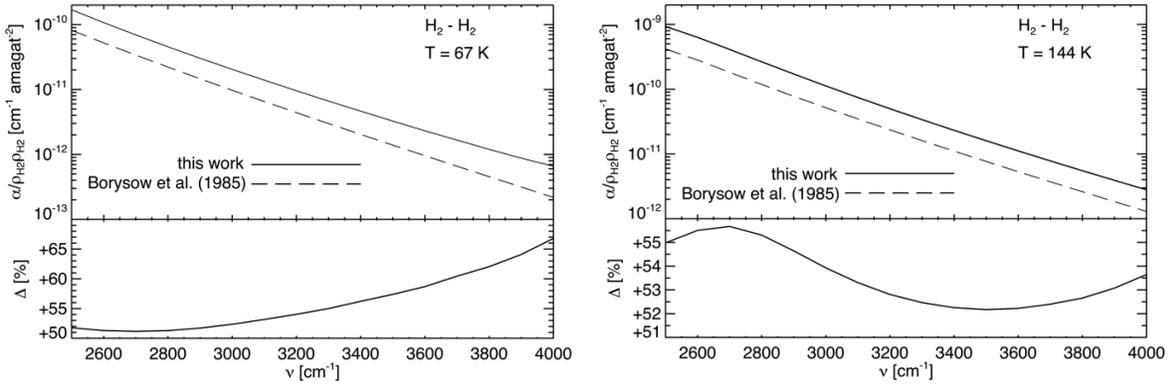

figure 9. Left: (top) Absorption coefficient spectra for the rototranslational $H_2$-$H_2$ collision-induced spectra at temperatures close to the mean brightness temperature of the upwelling radiance relevant to Uranus from Borysow et al.[27] (dashed lines) and for this work (solid lines). (bottom) Percent difference between the coefficients. Right: Similar to the left panel, but for a temperature relevant to Jupiter. The temperatures chosen for this illustration are the same as those in Fig. 3.

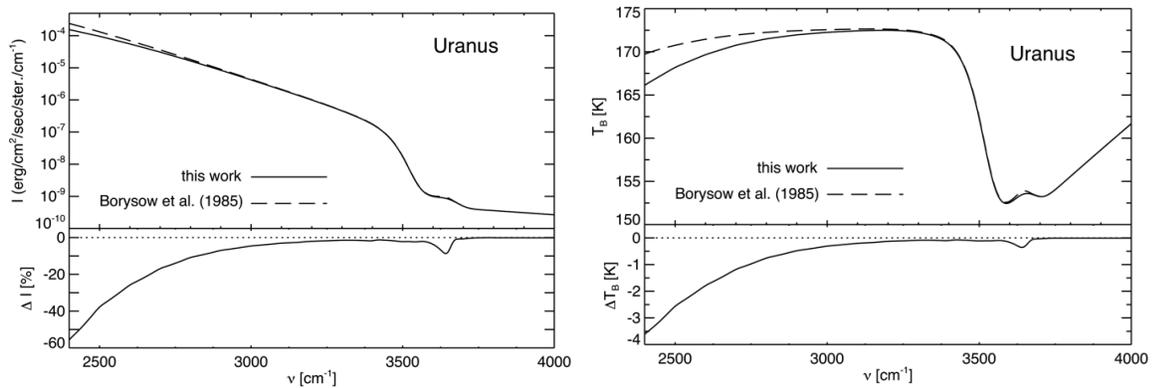

Figure. 10. Left: (top) Models for the radiance of nadir thermal emission emerging from Uranus, considering only the opacity provided by $H_2$-related collision-induced absorption, assuming equilibrium *para*-to-*ortho* fractions of $H_2$ at each level. The differences in the spectra result from using absorption coefficients for $H_2$-$H_2$ CIA based on Borysow et al.[27] (dashed line) and for those in this paper (solid line). (bottom) Spectrum of the percentage difference between radiance of the models. Right: Similar to the left panel but displaying the (top) radiance results in terms of equivalent brightness temperature and (bottom) brightness-temperature differences. The absorption at 3568 cm$^{-1}$ is from the fundamental O(3) transition.

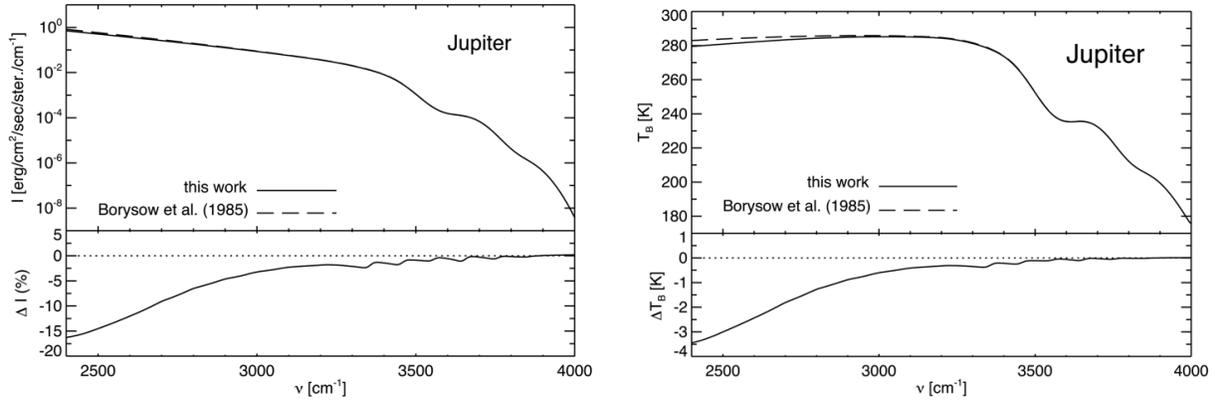

Figure 11. Similar to Fig. 10 but for the atmosphere of Jupiter. Additional absorptions from the 3807 cm$^{-1}$ O(2) and 3329 cm$^{-1}$ O(4) fundamental transitions are detectable.

## D. $H_2$-$H_2$, fundamental band

Figure 12 illustrates the full range of absorption coefficients for $H_2$-$H_2$ and $H_2$-He absorption that we have addressed in this work. This figure notes that absorption with arbitrary choices of $f_p$, the *para*-$H_2$ fraction between 25 and 100%, can be determined from 0 to 7400 cm$^{-1}$, encompassing both rototranslational and rotovibrational absorption. For $H_2$-$H_2$ collisions, this is the result of augmenting the FORTRAN code of Borysow[22], enabling coverage of this full spectral range. We note that we have not addressed differentiating between absorptions at different $f_p$ values for overtones and hot bands, which is why the values shown at wavenumbers greater than

7400 cm$^{-1}$ represent equilibrium $f_p$ values at each temperature.

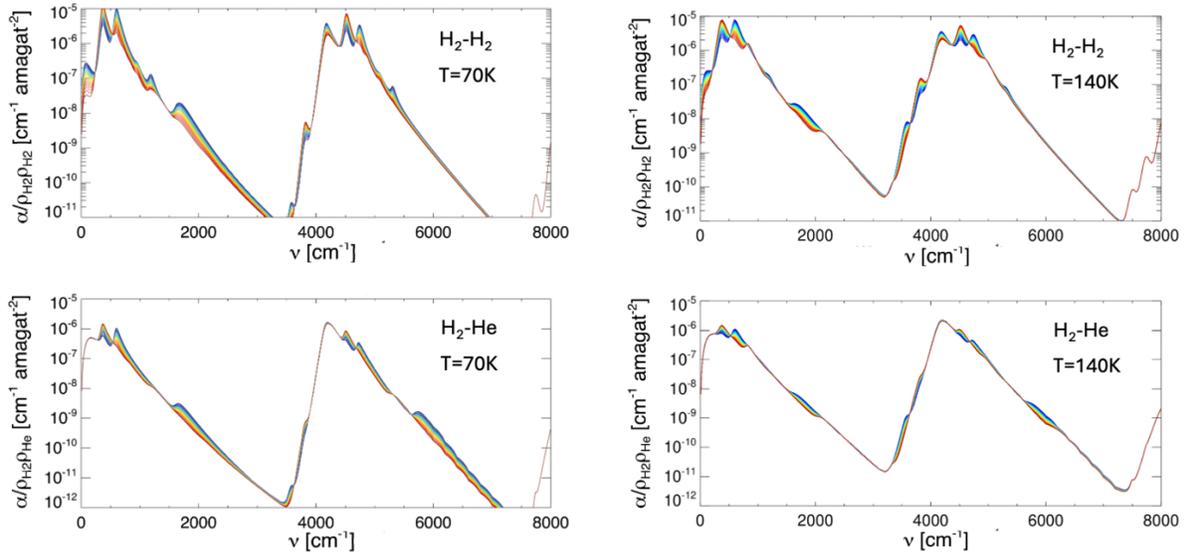

Figure 12. Absorption coefficients for H$_2$-H$_2$ and H$_2$-He covering 0 – 8000 cm$^{-1}$ at the temperatures designated and *para*-H$_2$ fractions, $f_p$, depicted by color, ranging from 25% (the minimum value expected at high temperatures) in blue through 100% in red. The temperatures chosen for this illustration are close to those in Figs. 4, 5, 7 and 9 and correspond directly to temperatures in the publicly available Zenodo file. Values for wavenumbers greater than 7400 cm$^{-1}$ are for equilibrium H$_2$.

**IV. Discussion / Conclusions**

We have created several updates of existing models and tabulated information for the collision-induced absorption of H$_2$-H$_2$ and H$_2$-He covering the spectral range of 0 – 8000 cm$^{-1}$. For the mid- and far-infrared, we have updated model spectra of H$_2$-He rototranslational collision-induced absorption, now based on *ab initio* calculations. These absorption coefficients cover the 40 - 400K temperature range. They have been extended from the shortest wavenumber of 20 cm$^{-1}$ up to 4000 cm$^{-1}$, providing a smooth merger with collision-induced rotovibrational absorption spectra of H$_2$-He in a region dominated by the fundamental-band absorption. For this higher spectral range, the H$_2$-He absorption has similarly been re-assessed using *ab initio* calculations. For completeness, for H$_2$-H$_2$ CIA absorption we have (1) extended the rototranslational models[2,3] out to 4000 cm$^{-1}$ and (2) addressed the fundamental-dominated absorption to the extent of generalizing the program of Borysow[22] to include a range of choices for the *para*-H$_2$ fraction.

The most substantial effects of the updated CIA coefficients appear in the near-infrared, where differences from current models can be well over 100% for H$_2$-He absorption between 5000 and 8000 cm$^{-1}$ and over 50% for H$_2$-H$_2$ rototranslational absorption between 2500 and 4000 cm$^{-1}$. For thermal emission in the 2400-4000 cm$^{-1}$ range, Figures 10 and 11 illustrate that our model for H$_2$-H$_2$ absorption would result in reduced emission, equivalent to a lower brightness temperature. Although the H$_2$-He absorption for the spectral region in our model is higher than the previous work by Borysow et al.[22] (see Fig. 4), the volume mixing ratio of He is nearly an order of magnitude lower than H$_2$, so the total change is dominated by the change in the H$_2$-H$_2$ absorption. In order to match the same observed thermal emission, then, radiative-transfer calculations using our models would need to increase the temperature assumed or reduce the atmospheric opacity, e.g. with reduced particulate absorption.

Figure 8 shows that at higher frequencies where the emerging radiation is dominated by reflected sunlight, the difference is more complicated. Differences in the depth of unit optical depth calculated from our models versus those derived by Borysow[24] are very small near the peaks of absorption by the fundamental band, sensitive to particulate reflection at and at altitudes above the 0.1-bar level. Further from the U-band peaks where the upwelling reflected sunlight is sensitive to particles as deep as several bars of atmospheric pressure, the differences in the atmosphere of Uranus are different from the atmosphere of Jupiter. Such differences could require adjustments in the vertical distribution of reflected particles: the particles must be moved to higher / deeper altitudes where our CIA coefficients reach unity at lower / higher pressures than those of Borysow[23]. The need for such changes in this region may be mitigated by the presence of strong absorption by vibration-rotation lines of CH$_4$, which is common in all giant-planet near-infrared spectra.

The effects of the updated coefficients in the rototranslational band covering the mid- to far-infrared are much smaller by comparison, but they should not be considered inconsequential. Of fundamental importance, our models have established confidence in the H$_2$-He CIA rototranslational band absorptions using a fundamental *ab initio* approach that is as well-grounded as for the H$_2$-H$_2$ CIA rototranslational spectrum[2,3]. Moreover, in their practical use, the changes our model makes in the H$_2$-He rototranslational spectrum, however subtle they may appear, have consequences in exploring the bulk compositional information to which this part of the spectrum is sensitive. To illustrate this, we conducted a simple test of the sensitivity of the spectrum to the He/H$_2$ ratio, finding that a fit to a noiseless spectrum over 50-550 cm$^{-1}$ generated by a model of Uranus with a [He]/[H$_2$] ratio by volume of 0.17 using our H$_2$-He absorption model was best fit by a spectrum generated using the Borysow et al.[24] $H_2$-He CIA model with an [He]/[$H_2$] ratio of 0.19. The 0.17 ratio is consistent with results derived by Voyager-2 comparisons of far-infrared observations and radio-occultation results[29], for which there is keen interest in verifying or improving because of the importance of knowledge of bulk composition to models for evolution and interior structure. Prior to a direct probe, the only method for determining this ratio is to fit observations of this part of the spectrum of Uranus, which is dominated by collision-induced absorption and is sensitive to the He/$H_2$ ratio. Use of our absorption models thus eliminates a potential +2% systematic bias in the bulk composition of Uranus using its spectrum alone. We note that this argument is also applicable to determining the bulk composition of Neptune's atmosphere.

The uncertainty associated with our absorption model depends on the interaction involved. The $H_2$-He rototranslational band uncertainty is estimated as being 2% up to 3000 cm$^{-1}$, essentially reflecting the numerical accuracy of the calculation. This becomes slightly worse at wavenumbers higher than 3000 cm$^{-1}$, rising to about 5% at 4000 cm$^{-1}$, where the rototranslational absorption is obscured by the fundamental absorption. We consider these as state-of-the-art and do not anticipate significant improvements in the near future. The $H_2$-$H_2$ rototranslational band uncertainty is estimated as 5%, a value derived from considerations of both the numerical accuracy and some uncertainty associated with the potential and dipole data. For this band, the isotropic potential approximation has turned out to give reliable results. So the absorption data used in this work (from references 1 and 2) are expected to be quite reliable. The $H_2$-He fundamental band uncertainty is estimated to be 15% at its worst, around 4600 cm$^{-1}$, and otherwise in the high absorption region (3800-5000 cm$^{-1}$) the accuracy is better. This is based on the discrepancy between the isotropic approximation and the full anisotropic calculation. These coefficients can be improved some, as they were computed using the isotropic potential approximation. The $H_2$-$H_2$ fundamental band uncertainty is estimated as 20%, based on the uncertainty associated with the available potential and dipole data. These coefficients can probably be improved more than the $H_2$-He coefficients in the fundamental, if better dipole and potential data were employed. We leave these recommended improvements in the fundamental-band absorption for future work.

For outer-solar-system radiative-transfer calculations we recommend use of our models in place of those of Abel et al.[17], which is currently used as a CIA standard in HITRAN absorption models in their high-temperature standard database.[28] Our calculations have been optimized for the temperatures of the giant planets, which are at the coldest limit of the range considered by Abel et al.[17], who focused their attention on the atmospheres of much hotter exoplanets. We note that the models we present here should also replace the HITRAN models addressing lower-temperature conditions (Refs. 22-24, 27) that are used in the HITRAN low-temperature auxiliary data base.

## V. Data Availability

All of the updated absorption coefficient files described in this paper can be obtained from: https://zenodo.org/doi/10.5281/zenodo.12687188. The 'README.md' file on this site identifies and describes each of them in some detail. The FORTRAN programs generating the coefficients in Borysow et al.[22] and Borysow[23] (shown in Figs. 3 and 6, respectively) are available from: https://www.astro.ku.dk/aborysow/programs/index.html. The coefficients have also been translated into a format that is suitable for use with the code used to generate the model spectra[29].

## CRediT authorships contribution statement

G. S. Orton: Writing – original draft, Visualization, Investigation, Funding acquisition. M. Gustafsson: Writing – original draft, Software, Investigation, Conceptualization, Funding acquisition. L. N. Fletcher: Writing – review & editing, Software, Conceptualization, Funding acquisition. M. T. Roman: Writing – review & editing, Software. J. A. Sinclair: Writing – review & editing, Software

## Declaration of competing interest

The authors declare that they have no known competing financial interests or personal relationships that could have appeared to influence the work reported in this paper.

## Acknowledgements

Research performed by G. S. Orton and J. A. Sinclair was carried out at the Jet Propulsion Laboratory, California Institute of Technology, under a contract with the National Aeronautics and Space Administration (80NM0018D0004). M. Gustafsson acknowledges support from the Knut and Alice Wallenberg Foundation. L. N. Fletcher and M. T. Roman were supported by STFC Consolidated Grant reference ST/W00089X/1. For the purpose of open access, the author has applied a Creative Commons Attribution (CC BY) license to the Author Accepted Manuscript version arising from this submission.

This research used the NEMESIS radiative-transfer code[30] to generate the spectral models displayed in Figs. 4, 5, 7, 8 and 10, for which we converted the new model absorptions presented here into a format suitable for the NEMESIS code. These calculations were performed efficiently using (1) JPL's High Performance Computing (HPC) resources, which were provided by funding from the JPL Information and Technology Solutions Directorate and (2) the ALICE High Performance Computing Facility at the University of Leicester.

M. Gustafsson wishes to acknowledge the great inspiration received from his PhD supervisor, the late Lothar Frommhold (1930–2021), who introduced him to the subject of interaction induced spectroscopy. His contribution to the spectral line shape community cannot be overestimated. All the authors wish to acknowledge the pioneering work done in this field by Aleksandra Borysow that enabled research involving the analysis of infrared remote sensing of outer-planet atmospheres to progress to a substantial state.

**Appendix A: Details of $H_2$–He IPA computations**

Table I. Input parameters for the $H_2$–He calculations with the isotropic potential approximation in the rototranslational (RT) and fundamental (RV) bands. The number of dipole components is 5 for all temperatures; $\lambda L$ = 01, 21, 23, 43, 45 (see Gustafsson et al.[7]).

| $T(K)$ | spectral range (cm$^{-1}$) | $l_{max}$ | $j_{max}$ | $j'_{max}$ | $R_{min}$ (bohr) | $R_{max}$ (bohr) | $\Delta R$ (bohr) |
|---|---|---|---|---|---|---|---|
| 40.000 | 0.1 – 4000. RT | 14 | 5 | 9 | 2.5 | 40.0 | 0.02 |
|  | 2000.- 8000. RV | 14 | 5 | 9 | 2.5 | 40.0 | 0.02 |
| 51.662 | 0.1 – 2400. RT | 17 | 5 | 9 | 2.5 | 40.0 | 0.02 |
|  | 2500.- 4000. RT | 17 | 5 | 9 | 2.5 | 40.0 | 0.01 |
|  | 2000. – 8000. RV | 17 | 5 | 9 | 2.5 | 40.0 | 0.02 |
| 66.724 | 0.1 – 4000. RT | 19 | 5 | 9 | 2.5 | 40.0 | 0.02 |
|  | 2000.- 8000. RV | 19 | 5 | 9 | 2.5 | 40.0 | 0.015 |
| 86.177 | 0.1 – 4000. RT | 21 | 5 | 9 | 2.5 | 40.0 | 0.02 |
|  | 2000.- 8000. RV | 21 | 5 | 9 | 2.5 | 40.0 | 0.02 |
| 111.302 | 0.1 – 4000. RT | 23 | 5 | 9 | 2.5 | 40.0 | 0.02 |
|  | 2000.- 8000. RV | 23 | 5 | 9 | 2.5 | 40.0 | 0.02 |
| 143.753 | 0.1 – 2400. RT | 25 | 5 | 9 | 2.5 | 40.0 | 0.02 |
|  | 2500.- 4000. RT | 25 | 5 | 9 | 2.5 | 40.0 | 0.01 |
|  | 2000. – 8000. RV | 25 | 5 | 9 | 2.5 | 40.0 | 0.02 |
| 185.664 | 0.1 – 2400. RT | 27 | 5 | 9 | 2.5 | 40.0 | 0.02 |
| | 2500.- 4000. RT | 27 | 5 | 9 | 2.5 | 40.0 | 0.01 |
| | 2000. – 8000. RV | 27 | 5 | 9 | 2.5 | 40.0 | 0.01 |
| 239.794 | 0.1 – 4000. RT | 30 | 5 | 9 | 2.5 | 32.0 | 0.015 |
| | 2000.- 8000. RV | 30 | 5 | 9 | 2.5 | 32.0 | 0.015 |
| 309.705 | 0.1 – 4000. RT | 32 | 5 | 9 | 2.5 | 32.0 | 0.015 |
| | 2000.- 8000. RV | 32 | 5 | 9 | 2.5 | 32.0 | 0.015 |
| 400.000 | 0.1 – 4000. RT | 34 | 5 | 9 | 2.5 | 25.0 | 0.015 |
| | 2000.- 8000. RV | 34 | 5 | 9 | 2.5 | 25.0 | 0.015 |

In this work the IPA calculations[9], in both the rototranslational and the fundamental band, were carried out with an in-house implementation of the Numerov algorithm for integration of the radial Schrödinger equation. A separate set of parameter values were used at each of the temperatures. Those are listed in Table I. A partial wave expansion is applied where $l_{max}$ is the highest end-over-end angular momentum quantum number. The $H_2$ rotational quantum numbers are allowed to vary from 0 up to $j_{max}$ and $j'_{max}$ for initial and final states, respectively. The largest value of $\lambda$ included in the expansion of the induced dipole moment is 4, and as a consequence $j'_{max}$ is $j_{max}+4$. Note that the even-j calculations can be carried out separately from the odd-j calculations, since there is no conversion between *para-* and *ortho*-$H_2$ (in absence of magnetic materials). Naturally, this holds for both IPA and CC calculations described below in Appendix B. The Numerov integration runs on a grid from $R_{min}$ to $R_{max}$ in steps of $\Delta R$. In order to energy normalize the continuum wave function, the integration continues past $R_{max}$ so that the phase shift can be determined. Our phase-shift search algorithm is sensitive to the step size, $\Delta R$, and that is why those values in Table I seem oddly small. The calculations are still significantly less elaborate than the CC calculations, so we can live with the small steps. At a few temperatures in the rototranslational calculation (52 K, 144 K, and 186 K) there is a need to decrease $\Delta R$ at the highest frequencies. This is due to the fast oscillations in the continuum wave function, which results from the energy deposited into the molecular system when it absorbs the photon.

**Appendix B: Details of $H_2$–He CC computations in the rototranslational band**

Figure 13 shows the CC calculations using the BSP potential[11] and the SK potential[13], which are the potentials of choice in this work and in Gustafsson et al.[7], respectively, at two of our tabulated temperatures, 86 K and 186 K. The more accurate, i.e. the BSP, potential gives an enhancement of the absorption. We saw in section IIA that the CC calculations of this work gives excellent agreement with laboratory measurements, which is an improvement over the previous calculation[6]. There is a development of the BSP potential energy surface, BSP3[32], that describes vibrationally excited $H_2$ better. In the calculations described here, $H_2$ remains in the vibrational ground state and we expect no discernible difference if the potential from Thibault et al.[32] would be used. We have performed an isotropic test calculation at 144 K that showed an overall difference of around, or less than, 1% between the absorptions computed with BSP and BSP3. However, for fundamental band calculations, and for vibrational overtones, one should consider using the BSP3 potential. The close-coupling program COUPLE[12] comes with both a Gordon[34] and a Numerov[34,35] integrator. The user can chose which one to use with an input switch. Both have been used in this work. For most data points both integrators work well; and having result from both gives us a chance to verify agreement between them. For some data points the Gordon integrator is unstable, so for those we have used the Numerov result.

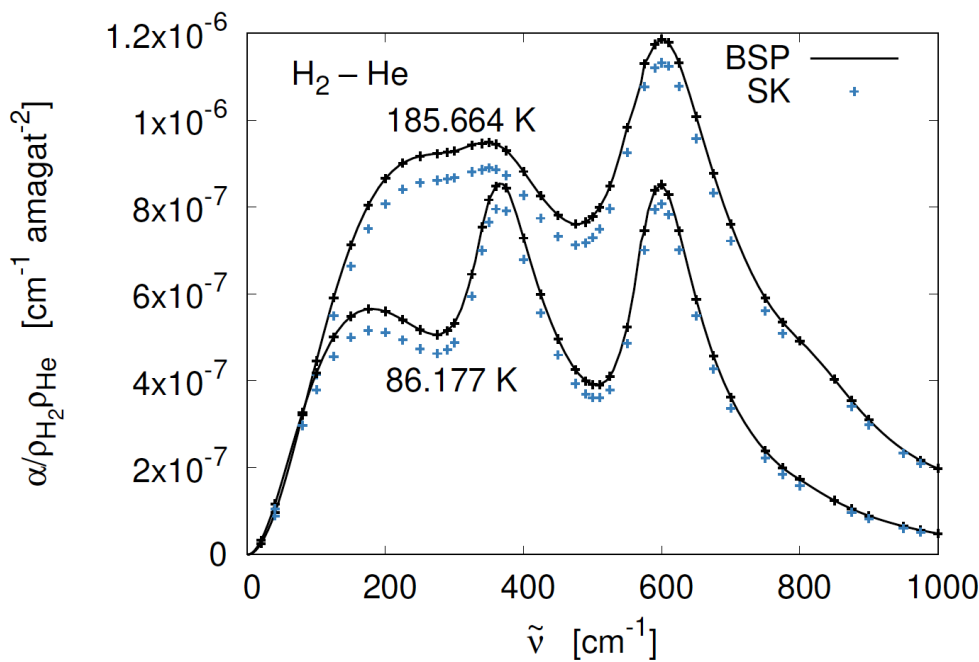

Figure 13. Binary collision-induced $H_2$–He absorption coefficient, α, normalized by the hydrogen and helium gas densities, as function of frequency in the rototranslational band, at the temperatures of 186 K (upper trace) and 86 K (lower trace). Equilibrium hydrogen, with respect to *para*-to-*ortho* distribution, is considered at both temperatures. All data are computed with the close coupling approach[7], which accounts for the full anisotropic potential. The blue symbols (SK) are computed with the ab initio potential surface[32] as given Schäfer and Köhler[13] which is the potential that was also used by Gustafsson et al.[7] The black symbols (BSP) are computed with the Bakr-

Smith-Patkowski potential energy surface[11], which is the one used for rototranslational $H_2$–He absorption in this work. The solid black curves represent the amended anisotropic BSP data, obtained with the AA method outlined in Section II A.

The parameter settings are given in Table II. The Gordon integrator works with an adaptive step in the radial coordinate; as a consequence no parameter $\Delta R$ is given for those calculations. For the high spectral region (3000 – 4000 cm$^{-1}$) the Gordon integrator becomes unstable. We think that this is because of the diversity in kinetic energies for the different channels. Say there is approach in the j = 1 channel, and a 4000 cm$^{-1}$ photon is absorbed, giving rise to a rotational transition j = 5, which is supported by the dipole components with $\lambda = 4$. The rotational transition then requires 1622 cm$^{-1}$ and there is an added kinetic energy to the system of almost 2400 cm$^{-1}$. It appears that handling channels of vastly different kinetic energy is problematic in the implementation of the Gordon integrator in COUPLE. Thus, we have used the Numerov results in the high spectral region. Our rather large value of $R_{max}$, which is 120 bohrs provides good stability for both integrators according to our tests. Most of our calculations did finish and give converged results even with $R_{max}$ set to a smaller value, but for some frequencies numerical problems appeared. Presumably the difficulty lies in the application of the asymptotic boundary conditions.

Table II. Input parameters for the H$_2$-He close coupling calculations in the rototranslational band. The number of potential components is three at all temperatures, γ = 0, 2, 4. The number of dipole components is give as in the isotropic calculations, see Table 1.

| Method | T(K) | $J_{max}$ | $j_{max}$ | $j'_{max}$ | $R_{min}$ (bohr) | $R_{max}$ (bohr) | $\Delta R$ (bohr) |
|---|---|---|---|---|---|---|---|
| Numerov | 40.000 | 17 | 5 | 7 | 2 | 120 | 0.1 |
| Gordon | 40.000 | 17 | 5 | 7 | 2 | 120 | - |
| Numerov | 51.662 | 20 | 5 | 7 | 2 | 120 | 0.1 |
| Gordon | 51.662 | 20 | 5 | 7 | 2 | 120 | - |
| Numerov | 66.724 | 22 | 5 | 7 | 2 | 120 | 0.1 |
| Gordon | 66.724 | 22 | 5 | 7 | 2 | 120 | - |
| Numerov | 86.177 | 24 | 5 | 7 | 2 | 120 | 0.1 |
| Gordon | 86.177 | 24 | 5 | 7 | 2 | 120 | - |
| Numerov | 111.302 | 26 | 5 | 7 | 2 | 120 | 0.1 |
| Gordon | 111.302 | 26 | 5 | 7 | 2 | 120 | - |
| Numerov | 143.753 | 28 | 5 | 7 | 2 | 120 | 0.1 |
| Gordon | 143.753 | 28 | 5 | 7 | 2 | 120 | - |
| Numerov | 185.664 | 30 | 5 | 7 | 2 | 120 | 0.1 |
| Gordon | 186.664 | 30 | 5 | 7 | 2 | 120 | - |
| Numerov | 239.794 | 33 | 5 | 7 | 2 | 120 | 0.1 |
| Gordon | 239.794 | 33 | 5 | 7 | 2 | 120 | - |
| Numerov | 309.705 | 35 | 5 | 7 | 2 | 120 | 0.1 |
| Gordon | 309.705 | 35 | 5 | 7 | 2 | 120 | - |
| Numerov | 400.000 | 37 | 5 | 7 | 2 | 120 | 0.1 |
| Gordon | 400.000 | 37 | 5 | 7 | 2 | 120 | - |

The Numerov algorithm operates with a predictor-corrector scheme, and in the COUPLE implementation it worked well with a step of 0.1 bohrs over the whole spectral range, as indicated in Table II. We tested with a smaller step to verify this. At 51.662 K the absorption coefficient is accurate within 5% at the highest energy 4000 cm−1 and well within 1% up to 3000 cm$^{-1}$. We accept a higher numerical error at 4000 cm$^{-1}$ as this lies right under the highly absorbing H$_2$ fundamental band.

It is necessary to include closed channels, i.e. channels that have energy below their asymptotic energy, in the CC calculations. We set a cutoff energy for inclusion of closed channels. Typically a value of 1097 cm−1 has been sufficient for convergence. However, for the highest frequencies, 3500 cm$^{-1}$ and 4000 cm$^{-1}$ that value did not provide convergence. We set the cutoff to 1622 cm$^{-1}$ for all calculations $\geq$ 2500 cm$^{-1}$ to ensure that $j = 5$ (the highest initial rotation) was included from the outset.

Another observation in the high frequency region is that the U(4) and U(5) transitions, at about 2885 cm$^{-1}$ and 3266 cm$^{-1}$, respectively, are visible in the IPA calculation for temperatures 186 K, 240 K, 310 K, and 400 K. Those transitions are, however, barely discernible for the lowest of those four temperatures. The CC code is not set up to include those transitions, so in order to account for them we have excluded CC data points at 3000 cm$^{-1}$ and 3500 cm$^{-1}$ for the four temperatures 186 K, 240 K, 310 K, and 400 K. This admittedly gives a quite sparse grid to determine amended anisotropic data according to the method outlined in Section II A. However, we accept somewhat larger uncertainty in our data in this frequency region of the rototranslational band, as it is largely obscured by the fundamental band absorption, in particular at T > 186 K.

The dipole coupling in the CC scheme includes the photon field intensity as a parameter[8,36]. This allows for non-linear (e.g. two-photon absorption) if the field strength is sufficiently high. Here, on the other hand, we want the process to be conventional one-photon absorption. We have performed a test to ensure that the absorption coefficient is independent of the field intensity, verifying that it is indeed one-photon absorption[6] we compute.

**Appendix C: Computational details for H$_2$–H$_2$ in the rototranslational band**

The parameter values for the calculations of the rototranslational H$_2$–H$_2$ collision-induced absorption are listed in Table III. These calculations are carried out in the isotropic potential approximation using our in-house Numerov integrator. The partial wave expansion goes from 0 to $l_{max}$, where $l$ is the end-over-end angular momentum quantum number. The H$_2$ rotational quantum numbers are indicated $j_i$ where $i = 1, 2$ for each of the two molecules. Those rotations run from 0 to $j_{i,max}$ and $j'_{i,max}$ for the initial and final states, respectively. The radial grid, over which the radial wave function is integrated, is defined by $R_{min}$, $R_{max}$, and $\Delta R$.

Table III. Input parameters for the $H_2$-$H_2$ calculations in the spectral region 2500 – 4000 cm$^{-1}$ of the rototranslational band. The number of dipole components is 8 for all temperatures, $\lambda_1\lambda_2\lambda L$ = 0001, 0221, 2021, 0223, 2023, 2233, 0445, 4045.

| T(K) | $l_{max}$ | $j_{i,max}$ | $j'_{i,max}$ | $R_{min}$ (bohr) | $R_{max}$ (bohr) | $\Delta R$ (bohr) |
|---|---|---|---|---|---|---|
| 40.000 | 14 | 1 | 7 | 3.0 | 40.0 | 0.025 |
| 51.662 | 17 | 1 | 7 | 3.0 | 40.0 | 0.025 |
| 66.724 | 19 | 2 | 6 | 3.0 | 40.0 | 0.020 |
| 86.177 | 21 | 2 | 6 | 3.0 | 40.0 | 0.020 |
| 111.302 | 23 | 3 | 7 | 3.0 | 40.0 | 0.020 |
| 143.753 | 25 | 3 | 7 | 3.0 | 40.0 | 0.020 |
| 185.664 | 27 | 4 | 8 | 3.0 | 40.0 | 0.020 |
| 239.794 | 30 | 5 | 9 | 3.0 | 32.0 | 0.015 |
| 309.705 | 32 | 5 | 9 | 3.0 | 32.0 | 0.015 |
| 400.00 | 34 | 5 | 9 | 3.0 | 25.0 | 0.015 |

**Appendix D: Absorption coefficient with equilibrium versus arbitrary *para/ortho*-fraction**

Here the formulas for calculation of the collision-induced absorption coefficient for equilibrium $H_2$ are reviewed, and formulas for an arbitrary *para*- $H_2$ ratio, $f_p$, are presented. This way we attempt to make clear exactly what differences there are in the statistical physics description of the two cases. The distinction is useful since *para-ortho* conversion only happens when there is magnetic materials present, or if chemical reactions break and reform the bonds in the $H_2$ molecules. So, if hydrogen gas is cooled or heated, in absence of the above conversion mechanisms, it will end up out of equilibrium from a nuclear spin point of view. The *para*-$H_2$ and ortho-$H_2$ molecules correspond to nuclear spin singlet and triplet states, respectively.

1. $H_2$–He

The statistical weight for rotational state j in equilibrium $H_2$ is

$$\omega_j(T) = \frac{g_j(2j+1)e^{-E_j/k_B T}}{\sum_{j'} g_{j'} (2j'+1)e^{-E_{j'}/k_B T}} \quad (D1)$$

where $g_j$ is 1 and 3 for *para*-$H_2$ (even j) and ortho-$H_2$ (odd j), respectively. $E_j$ is the rotational energy, $k_B$ is Boltzmann's constant and T is the absolute temperature. The equilibrium absorption coefficient can then be written

$$\alpha(T,\tilde{v}) = \sum_j \omega_j(T)\alpha_j(T,\tilde{v}) \tag{D2}$$

where $\alpha_j(T,\tilde{v})$ is the *j*-resolved absorption coefficient, which can be obtained by identifying, and excluding $\sum_j \omega_j(T)$ in the expression for the absorption coefficient in the IPA or CC formulation in Birnbaum et al.[9] or Gustafsson et al.[7], respectively.

The statistical weights for arbitrary *para/ortho*-$H_2$ fraction, i.e. non-equilibrium $H_2$ gas, are

$$\omega_j^p(T,f_p) = f_p \frac{(2j+1)e^{-E_j/k_B T}}{\sum_{even\,j'}(2j'+1)e^{-E_{j'}/k_B T}}$$

$$\omega_j^o(T,f_p) = (1-f_p) \frac{(2j+1)e^{-E_j/k_B T}}{\sum_{odd\,j'}(2j'+1)e^{-E_{j'}/k_B T}} \tag{D3}$$

where $f_p$ is the partial abundance of *para*-$H_2$, which is a number between 0 and 1, so that the partial abundance of ortho-$H_2$ is $1-f_p$. The formulas (D3) express that, among the molecules with even j, there is rotational equilibrium, as well as there is among the molecules with odd j. But, there is in general not equilibrium between the even and odd j populations. The absorption coefficient in this case is

$$\alpha(T,\tilde{v},f_p) = \sum_{even\,j} \omega_j^p(T,f_p)\alpha_j(T,\tilde{v}) + \sum_{odd\,j} \omega_j^o(T,f_p)\alpha_j(T,\tilde{v}) \tag{D4}$$

where $\alpha_j(T,\tilde{v})$ has the same meaning as in Eq. (D2).

2. $H_2$–$H_2$

The definitions of the weights in Eqs. (D1) and (D3) are convenient to use in the $H_2$–$H_2$ case as well. The equilibrium absorption coefficient is

$$\alpha(T,\tilde{v}) = \sum_{j_1 j_2} \omega_{j_1}(T)\omega_{j_2}(T)\alpha_{j_1 j_2}(T,\tilde{v}) \tag{D5}$$

where again, the $H_2$ rotation dependent absorption coefficient $\alpha_{j_1 j_2}(T,\tilde{v})$ can be identified in formulas given by Meyer et al.[21] for the IPA case. $H_2$–$H_2$ is not treated with the CC scheme in this work but if needed, then formulas are given in Gustafsson et al.[37] The non-equilibrium, $f_p$-dependent, absorption coefficient is

$$\alpha(T,\tilde{v},f_p) = \sum_{even\,j_1}\sum_{even\,j_2} \omega_{j_1}^p(T,f_p)\omega_{j_2}^p(T,f_p)\alpha_{j_1 j_2}(T,\tilde{v})$$

$$+ \sum_{even j_1} \sum_{odd j_2} \omega_{j_1}^p(T, f_p) \, \omega_{j_2}^o(T, f_p) \alpha_{j_1 j_2}(T, \tilde{v})$$

$$+ \sum_{odd j_1} \sum_{even j_2} \omega_{j_1}^o(T, f_p) \, \omega_{j_2}^p(T, f_p) \alpha_{j_1 j_2}(T, \tilde{v})$$

$$+ \sum_{odd j_1} \sum_{odd j_2} \omega_{j_1}^o(T, f_p) \, \omega_{j_2}^o(T, f_p) \alpha_{j_1 j_2}(T, \tilde{v}) \tag{D6}$$